\documentclass[prl,reprint,showpacs]{revtex4-1}
\usepackage{times,amsmath,graphicx,latexsym}
\begin{document}

\title{Comment on ``Vortex String Formation in a 3D U(1) Temperature Quench"}

\maketitle

A computer simulation of the "formation" of vortex strings in variable-rate temperature quenches of a 3D XY superfluid through the ordering phase transition was carried out a number of years ago by Bettencourt \textit{et al.\,}\cite{zurek}, but the results found in those studies for the vortex line density as a function of time were never published, even in follow-up articles \cite{zurek2} on the same study.  None of the plots in Ref.\,1 could have been created without detailed knowledge of the time and quench rate dependence of the density.  The authors seek to test the Kibble-Zurek Mechanism (KZM) \cite{kz}, a scaling theory that predicts the vortex density at a particular ``freeze-out" time, a variable in the KZM  related to the point in the quench where the system no longer evolves adiabatically.  However, in the experiments in quenched superfluid $^4$He \cite{exp} it is the direct time dependence of the vortex density that is the main variable of interest, since the freeze-out time is not experimentally accessible. 

It has recently been shown in studies of the dynamics of 2D XY superfluids \cite{exact} that no vortex pairs are ``formed" in rapid quenches, there is only monotonic decay of the thermal vortices existing at the initial high temperature. Quenches at different rates showed that the fastest quenches in fact had the lowest density of vortices left over at all times following the start of the quench. These 2D results are the direct opposite of the predictions of KZM, which would claim that the highest density of vortices would be created for the fastest quench rates. 

We disagree with the claim in \cite{zurek} that ``Above $T_c$ \ldots strings are little more than nonperturbative field fluctuations.  Below $T_c$, they gradually acquire stability and can be regarded as \ldots vortex lines".  We assert that vortices remain well-defined at all times even above $T_c$, since even though the scale-dependent superfluid density is driven there to zero at long length scales, at short scales it remains non-zero.  Quantization of circulation thus remains valid at short scales, and hence vortex cores can still easily be located above $T_c$.  Previous 3D simulations of the XY model have had no trouble accessing the line density above $T_c$, and indeed Fig.\,1a of \cite{zurek} shows that their simulation is also easily able to measure the vortex density at all temperatures.  There is no reason then to assert (as in KZM) that the vortex density should only be measured at the freeze-out time.  Since the freeze-out time increases as a power of the quench time, this guarantees that fewer vortices will be ``formed" in slow quenches, since the density will be sampled only after very long times.  We very much doubt that that vortices are actually ``created" in a 3D quench: there will only be decay of the vortex loops thermally excited at the initial temperature, similar to the 2D case (and which is actually already evident in Fig.\,1a of \cite{zurek}).  

We do not particularly disagree with the KZM scaling method in general, but for the case of XY superfluids it does not appear to correctly account for the behavior of the thermal vortices both above $T_c$ and in the Ginzburg regime below $T_c$.  What it misses in slower quenches is the long-time survival of these thermal vortices, since the temperature remains high for a considerable time in a slow quench.  In the 2D studies \cite{exact} this effect was found to dominate the quench dynamics at intermediate times, with the curves for different quench rates only merging when the quench temperature reached its lowest value.  We speculate that the 3D data will be quite similar to the 2D results, with the fastest quench actually showing the lowest vortex density at all times since it most rapidly gets to the lowest temperature.

Without the vortex density data it is not possible to definitively test the main KZM prediction for the vortex density at the freeze-out time, Eq.\,8 of \cite{zurek}.  Such a plot is missing from the paper, even though the freeze-out time is displayed in their Fig.\,2.  Instead, the authors only display in Fig.\,3 the density at a completely different time, the very much longer time for the order parameter to become nearly equal to one, which is basically the time for the completion of the quench to $T$ = 0.  It is not at all clear why this time was substituted for the freeze-out time, since it was never mentioned in any previous KZM papers.

We would ask the authors of \cite{zurek} to take this opportunity to make public their data on the 3D vortex line density as a function of time and the quench rate.  Even a qualitative description of the time dependence of the data at different quench rates would be of help to the experimentalists.  We also point out that for any future 3D XY simulations it would be quite important to compute the recovery of the scale-dependent superfluid density as a function of both time and length scales.  Similar to the behavior for the 2D case seen in \cite{exact}, it would be expected that the smallest vortex loops would decay first, causing the initial recovery of the superfluid density only at the shortest scales.  With increasing time the recovery would expand to longer scales, but only at very long times would it reach the scales needed for experimental measurements (e.g.\ macroscopic second sound wavelengths \cite{exp}).
\linebreak

\noindent Gary A. Williams\\
Department of Physics and Astronomy\\
University of California, Los Angeles\\
Los Angeles, CA 90095\\

\end{document}